\documentclass[%
reprint,
groupedaddress,
amsmath,amssymb,
aps,
floatfix
]{revtex4-2}

\usepackage{graphicx}
\usepackage{siunitx}
\usepackage{dcolumn}
\usepackage{bm}
\usepackage{kantlipsum}
\usepackage{hyperref}
\usepackage[mathlines]{lineno}
\usepackage{cleveref}
\usepackage{physics}
\usepackage{subfiles}

\DeclareSIUnit\sample{Sa}
\begin{document}
\preprint{AIP/123-QED}
\title{Deterministic quantum phase estimation beyond the ideal NOON state limit}
\author{Jens A.\,H. Nielsen}
\author{Jonas S. Neergaard-Nielsen}
\author{Tobias Gehring}
\email{tobias.gehring@fysik.dtu.dk}
\author{Ulrik L. Andersen}
\email{ulrik.andersen@fysik.dtu.dk}
\affiliation{Center for Macroscopic Quantum States (bigQ), Department of Physics, Technical University of Denmark, Fysikvej, 2800 Kgs. Lyngby, Denmark}
\date{\today}

\begin{abstract}
The measurement of physical parameters is one of the main pillars of science. A classic example is the measurement of the optical phase enabled by optical interferometry where the best sensitivity achievable with $N$ photons scales as $1/N$ -- known as the Heisenberg limit~\cite{Helstrom1969,Caves1981,Braunstein1994,Giovannetti2011,Demkowicz-Dobrzanski2015,Polino2020}. To achieve phase estimation at the Heisenberg limit, it has been common to consider protocols based on highly complex NOON states of light~\cite{Dowling2008}. However, despite decades of research and several experimental explorations~\cite{Mitchell2004,Walther2004,Nagata2007,Daryanoosh2018,Slussarenko2017,Gao2010,Wang2016}, there has been no demonstration of deterministic phase estimation with NOON states reaching the Heisenberg limit or even surpassing the shot noise limit. Here we use a phase estimation scheme based on a deterministic source of Gaussian squeezed vacuum states and high-efficiency homodyne detection to obtain phase estimates with an extreme sensitivity that significantly surpasses the shot noise limit and even beats the performance of an ideal, and thus unrealistic, NOON state protocol. Using a high-efficiency setup with a total loss of about 11\% we achieve a Fisher Information of \SI{15.8 \pm 0.6}{\per\radian\squared}  
per photon unparalleled by any other optical phase estimation technology. The work represents a fundamental achievement in quantum metrology, and it opens the door to future quantum sensing technologies for the interrogation of light-sensitive biological systems~\cite{Taylor2016}.
\end{abstract}

\maketitle

\begin{figure*}[htp]
    \centering
    \includegraphics[width = \linewidth]{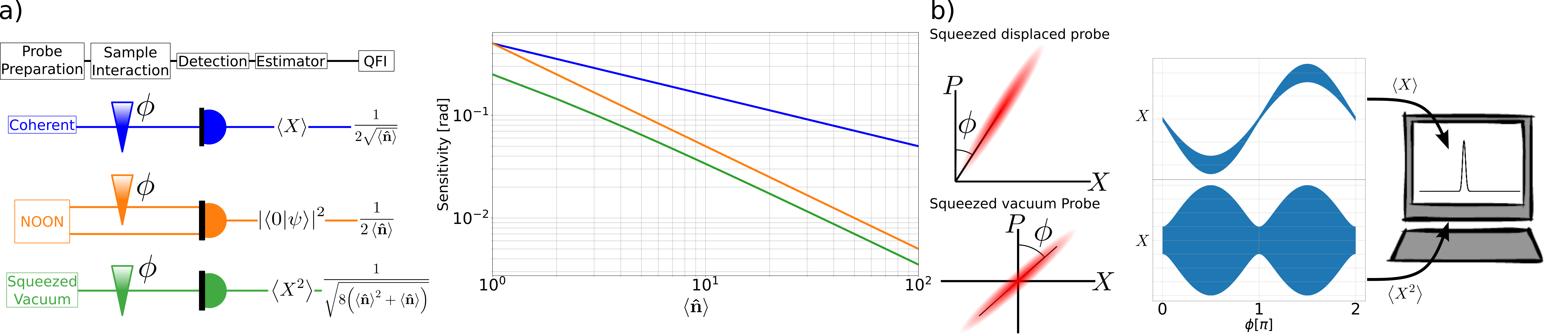}
    \caption{Principles and limits of quantum phase estimation. a) Schematics of three different phase estimation schemes. A quantum state of light undergoes a phase shift which is measured with either a homodyne detector (HD) or a NOON-state detector (involving photon counters) from which estimators are used to estimate the phase shift. Note that the NOON-state scheme is based on a two-beam interferometer in which only half of the photons traverse the sample. We therefore use the conservative sensitivity bound of $1/2\expval{\vu{n}}$ (where $N=2\expval{\vu{n}}$) for the comparison to our approach. b) The optimal sensitivities for the three schemes. c) Phase space pictures of a displaced squeezed state and a vacuum squeezed state, and the resulting quadrature measurements as a function of the phase. The phase is estimated using the estimators $\langle X\rangle$ or $\langle X^2\rangle$ for the displaced squeezed state and vacuum squeezed state, respectively.}
    \label{fig1}
\end{figure*}

It is of fundamental interest and practical relevance to investigate the ultimate bounds on the precision in estimating a phase~\cite{Helstrom1969,Braunstein1994}. According to classical (that is, approximate) theories of light, phase estimation can in principle be carried out with an arbitrary precision, but due to the inherent corpuscular quantum nature of light phase measurements will in reality be limited in precision – a precision that depends on the probing quantum state of light. If non-entangled states are  used, the ultimate precision limit is the shot-noise limit (SNL) where the sensitivity scales as $1/\sqrt{\langle \vu{n}\rangle}$, with $\langle \vu{n}\rangle$ being the average number of photons traversing the sample. By exploiting entangled states, it is possible to reach the ultimate Heisenberg limit with superior scaling (see \cref{fig1}a).

One of the most celebrated quantum states for reaching the ultimate Heisenberg limit -- often referred to as the optimal state for loss-free sensing -- is the so-called NOON state~\cite{Dowling2008}: $|\Psi_{NOON}\rangle=1/\sqrt{2}(|N\rangle|0\rangle +|0\rangle |N\rangle )$ which represents an optical state that is a superposition of $N$ photons across two optical modes. 
Although a large number of experimental realizations on phase estimation with NOON states have been reported~\cite{Mitchell2004,Walther2004,Nagata2007,Daryanoosh2018,Slussarenko2017,Gao2010,Wang2016}, as of today, only a single experiment has been able to obtain a sensitivity that violates the SNL~\cite{Slussarenko2017}, and even in this realization, the SNL was only beaten 
by using a probabilistic source of two-photon NOON states. Due to the high complexity in generating the NOON state and their extreme fragility to loss, it is unlikely that NOON states will be able to reach the Heisenberg limit, or even beat the SNL, for high photon numbers. 

It has been known for decades that the SNL can be more easily surpassed using squeezed states of light~\cite{Caves1981,Monras2006,Aspachs2009,Pinel2012}, which by now has also been realized in several phase estimation experiments~\cite{Grangier1987,Mckenzie2002,LIGO2013,Pradyumna2020,Lawrie2019}. However, in most of those experiments, squeezed light is combined with a bright coherent state in an interferometric measurement by which the sensitivity is 
often limited to $\sqrt{V_{-}/\langle \vu{n}\rangle}$ (where $V_{-}$ is the variance of the squeezed state quadrature normalized to the variance of the vacuum state). Although being superior to the shot noise limit of $1/\sqrt{\langle \vu{n}\rangle}$, the sensitivity is inferior to the Heisenberg scaling and thus does not reach the fundamental limit of NOON states. In fact, it is often claimed that Heisenberg scaling with squeezed light requires a highly complex measurement strategy~\cite{Anisimov2010}. However, in this Letter we show that by employing squeezed vacuum as a probe and a simple quadrature detector as the measurement device, phase estimates at the Heisenberg limit can be attained by evaluating the square of the quadrature outcomes. In addition to a sensitivity scaling similar to that for NOON states, our practical squeezed state estimation protocol is able to reach absolute sensitivities superior to those of the ideal NOON states due to a favorable scaling factor of $\sqrt{1/2}$~\cite{Monras2006}.  
We also note that in contrast to previous NOON state realizations, our scheme is not based on probabilistic sources of light or any post-selection of the measurement outcomes.  

\begin{figure*}[htp]
    \centering
    \includegraphics[width = \linewidth]{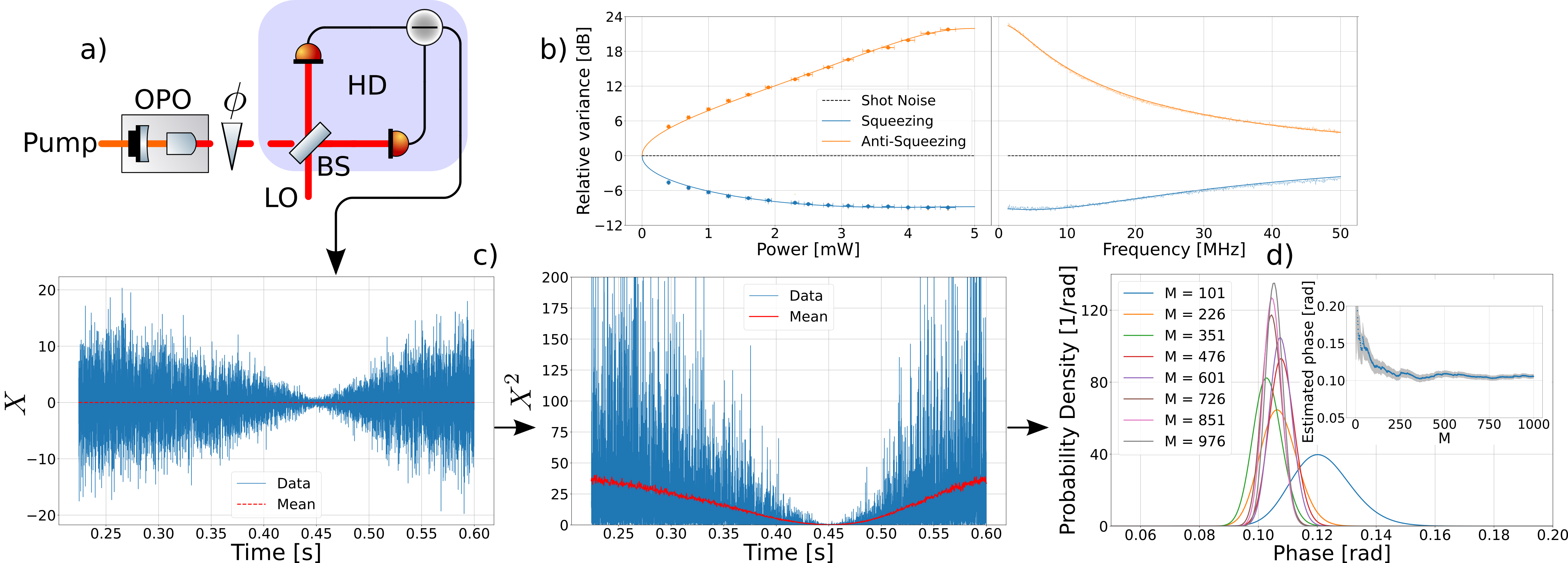}
    \caption{Experimental scheme and measurement method. a) Schematic of the experimental setup comprising an optical parametric oscillator (OPO) for squeezed light generation and a high-efficiency homodyne detector with a controllable local oscillator. As the estimated phase shift is relative between the squeezed vacuum and the local oscillator, in the experimental realization, we imposed the phase shift onto the local oscillator. b) Squeezed light spectrum and noise power versus pump power at the sideband frequency of 5 MHz for the squeezed and anti-squeezed quadratures. c) Quadrature measurement outcomes and their squares. The data are acquired while slowly varying the phase of the local oscillator, and down-mixed to a 5 MHz sideband frequency with a bandwidth of 1MHz. d) An example of a posteriori probabilities for the phase for different measurement trials and the associated phase estimates (inset).}
    \label{fig2}
\end{figure*}

\begin{figure*}[htp]
    \centering
    \includegraphics[width = \linewidth]{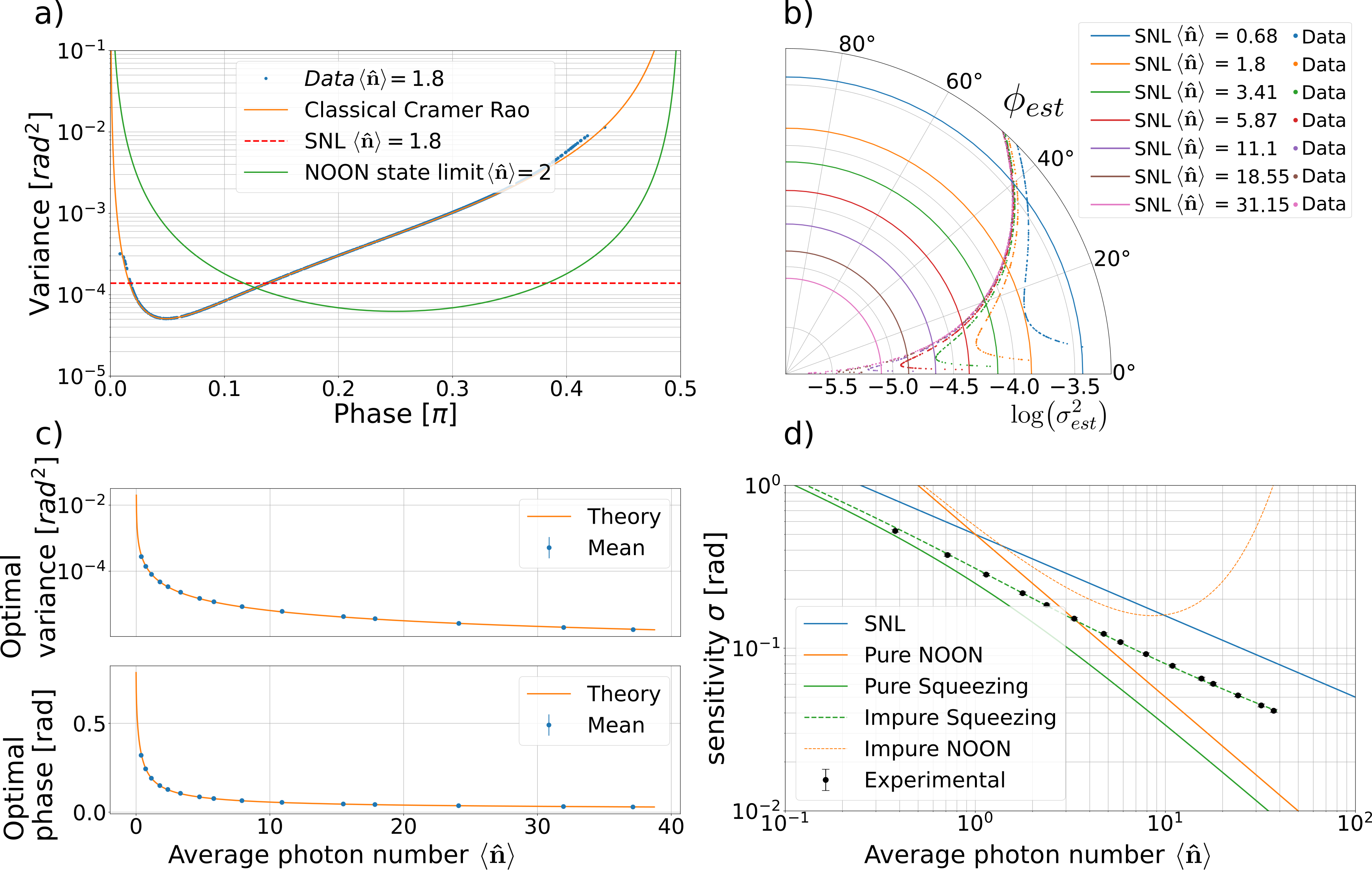}
    \caption{Quantum phase estimation results. a) The variance of the phase estimate based on 1000 quadrature measurements of a squeezed vacuum state with 11 photons. This is compared to the SNL and the limit for an ideal NOON state with $2\expval{\vu{n}} = N =4$. b) The variance of the phase estimated for different average photon numbers represented in a polar diagram and compared to the SNLs of the respective realizations (the curves are color-coded). It is clear from these plots that the variance is minimized for certain phases. The minimal variances and associated phases are presented in c) for different photon numbers and compared with theory. d) The optimized sensitivities versus photon numbers are presented and related to the theoretical predictions for the SNL, squeezed vacuum limit and the NOON state limit. We include theoretical predictions for the ideal limits and the practical limits with 11\% loss as measured in our system.}
    \label{fig3}
\end{figure*}
The conventional approach to squeezing-enhanced phase estimation is based on displaced squeezed states undergoing phase shifts that are estimated using a phase-referenced homodyne detector. The estimator, $X$, then yields an estimate of the phase with a quadrature uncertainty that depends on the actual phase as illustrated in \cref{fig1}b: The best phase estimate is achieved when the response (derivative of $\langle X\rangle$) is maximized and the noise is minimized which, in this case, occurs mid-fringe (at the phases $\phi=n\pi$ where $n=0,1,2…$). Using instead squeezed vacuum as the probe, the measurement of $X$ does not yield information about the phase since in this case $\langle X \rangle=0$, but if we use $X^2$ as the estimator, the information is revealed. In this case, however, the phase shift for which the response is the largest is not coinciding with the phases with minimum noise (at $\phi=n\pi$) and thus a trade-off needs to be found for which the sensitivity is optimized. The trade-off is optimized for the phases $\phi=\arccos(\tanh2r)/2+n\pi$ at which the Fisher Information is maximized; $F=2\sinh^2(2r)$ where $r$ is the squeezing parameter. From the Fisher Information, we find the sensitivity $\sigma_{sqz}=1/\sqrt{2\sinh^2(2r)}$ which can be expressed in terms of the average photon number (see supplementary material):
\begin{equation}
\sigma_{sqz}=\frac{1}{\sqrt{8(\expval{\vu{n}}^2+\expval{\vu{n}})}}\ .
\label{sigma}
\end{equation}
Here we assume a pure squeezed state; for impure squeezed states see the Supplementary Material. The expression in \cref{sigma} exhibits Heisenberg scaling (for $\expval{\vu{n}} \gg 1$), and moreover, it saturates the quantum Cramér--Rao bound which means that the scheme with homodyne detection of squeezed vacuum is optimal among all possible measurement strategies. In addition to being optimal among all Gaussians, it is also clear that it beats the complex estimation strategy of using non-Gaussian NOON states as $\sigma_{sqz}<1/2\expval{\vu{n}}$ for all $\expval{\vu{n}}$.

A simplified schematic of the experimental setup is shown in \cref{fig2}a (see Supplementary Material for details). We employ type 0 parametric down-conversion in a high-quality optical cavity to produce squeezed vacuum in a single spatial mode at the wavelength of 1550 nm. The squeezed vacuum state then experiences a phase shift of $\phi$ (relative to a reference) before its $X$ quadrature is measured by a homodyne detector. At this detector, the squeezed mode interferes with a phase-referenced local oscillator mode 
at a balanced beam splitter, the two outputs are detected with high-efficiency photodiodes, and the resulting currents are subtracted, amplified and fed to a computer for phase estimation
and analysis. 

By paying careful attention to the design and implementation of the source and the detectors, the total absorption and scattering loss was kept below $11\%$ including the loss associated with the source, the propagation and the detector. As a result, we produce squeezed states with a maximum of 9.0 dB of squeezing at a sideband frequency of 5 MHz (see \cref{fig2}b). Due to the absorption and scattering losses, the produced squeezed vacuum state is not pure, which eventually leads to a deviation from Heisenberg scaling of the sensitivity.
\begin{figure*}[htp]
    \centering
    \includegraphics[width = \linewidth]{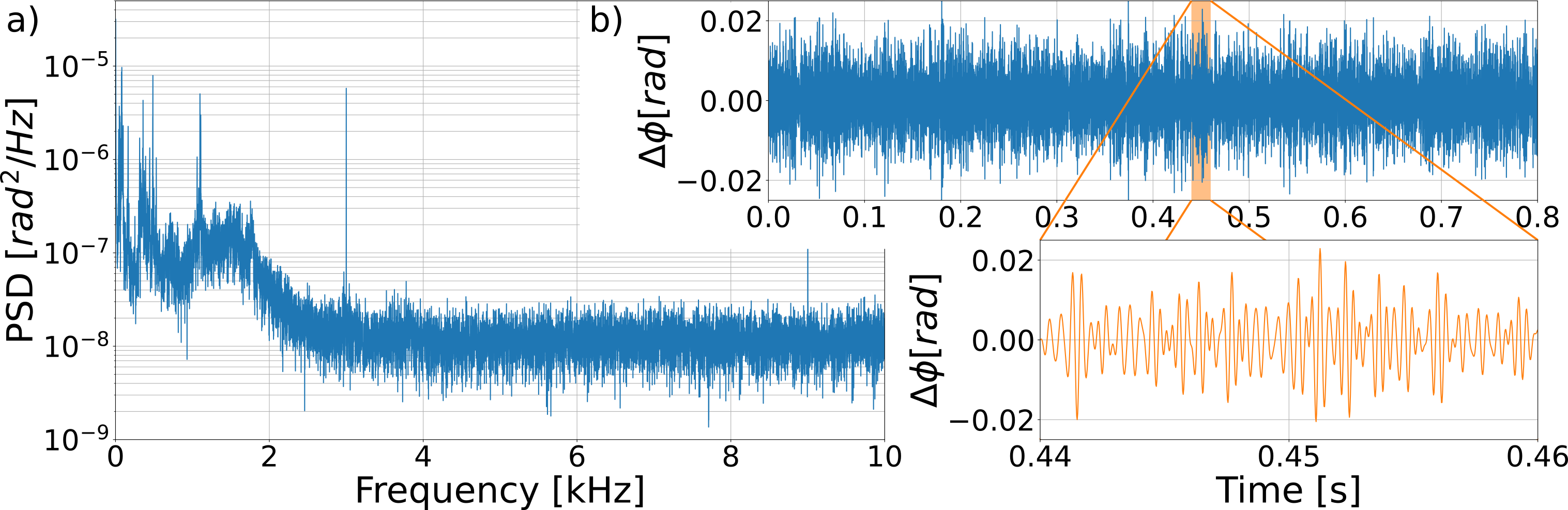}
    \caption{Quantum-enhanced tracking of a phase signal. a) Estimated dynamically varying phase signal using squeezed vacuum (with 6 photons) and Bayesian inference (inset) and the associated frequency spectrum. A 3 kHz induced signal as well as some low-frequency noise are apparent. b) Time trace of the same signal but bandpass-filtered at 3 kHz with a 2 kHz bandwidth. The zoom of the time trace as well as the frequency spectrum clearly shows the 3 kHz modulation.The y-axis $\Delta\phi$ is the relative phase shifts compared to the preset measurement phase.}
    \label{fig4}
\end{figure*}
To estimate the phase, $\phi$, and the associated uncertainty, we conduct $M=1000$ quadrature measurements for each phase setting, thereby producing a collection of 1000 data points, $\{x\}_M$. An example of the measured quadrature, $X$, and the conversion to $X^2$ for different phases are presented in \cref{fig2}c. From these measurements, we find the likelihood of acquiring the set $\{x\}_M$ conditioned on the phase $\phi$: $P(\{x\}_M|\phi)=\Pi_{i=1}^MP(x_i|\phi)$. The individual measurements are sampled from a Gaussian distribution, $P(x|\phi)=\exp\left(-x^2/2V(\phi)\right)/\sqrt{2\pi V(\phi)}$, 
with variance $V(\phi)=V_-\cos^2(\phi)+V_+\sin^2(\phi)$, where $V_+$ and $V_-$ are the anti-squeezed and squeezed variances, respectively. Using Bayes' theorem, we find the probability distribution for the phase conditioned on the measurement outcomes (the a posteriori probability distribution): $P(\phi|\{x\}_M)=P(\{x\}_M|\phi)P(\phi)/P(\{x\}_M)$ where $P(\{x\}_M)$ is a normalization factor and $P(\phi)=2/\phi$ is the a priori probability distribution of the phase (assumed to be flat in the range $[0;\pi/2]$). In \cref{fig2}d, we plot the a posteriori distribution for different values of M, illustrating the gradual Bayesian updating of the phase estimate. We then determine the estimated phase as the argument of the maximum value of $P(\phi|\{x\}_M)$ (see inset in \cref{fig2}d) and the associated phase uncertainty by the width of the distribution. These results are summarized in \cref{fig3}a for $\expval{\vu{n}}=1.8$ and in a polar plot representation in \cref{fig3}b for different average photon numbers. It is clear that the phase uncertainty decreases with increasing photon number (which we realize by varying the squeezing degree) and that it is optimized at specific phases (see \cref{fig3}c). The best operating principle of the system is thus to measure small phase shifts relative to the measurement angle for which the phase variance is smallest. In \cref{fig3}d, we plot the sensitivity optimized over the phase for different photon numbers, and we clearly observe performance beyond the ideal NOON state limit for photons up to around 3 as well as beyond the SNL and the loss-adapted NOON state limit for photons up to around 40. 




Since our states are being produced and measured deterministically, we are in a position to perform real-time measurements of a dynamically varying phase with near-ultimate precision. To do this, we probe an induced 3 kHz phase modulation as well as other low-frequency phase noise components with our sensitive probe which in these measurements contains 6 photons and preset (and locked with a bandwidth of less than approx 1kHz) to the optimal phase. The frequency spectrum of the measured phase signal and noise is shown in \cref{fig4}a and the real-time estimate of the dynamically varying phase is shown in \cref{fig4}b for $M=100$. By zooming into a certain time interval, the 3 kHz signal becomes visible (\cref{fig4}c). 

In summary, we have demonstrated phase sensing close to the ultimate limit, beating the ideal NOON state phase sensing scheme – often viewed as the optimal phase sensing strategy – with up to about 3 photons using solely squeezed vacuum and homodyne detection. To the best of our knowledge this is the best sensitivity per resource achieved in any optical phase sensing experiment: The directly measured Fisher Information per photon in our scheme is \SI{15.8\pm0.6}{\per\radian\squared} which should be contrasted to the Fisher information of the best NOON state experiment of $\sim\SI{4.2}{\per\radian\squared}$~\cite{Slussarenko2017}.
While we have demonstrated violations of the SNL and the NOON state limit for only a small range of phases, it can be easily extended to phases covering the entire range of $[0;\pi/2]$ by making use of an adaptive feedback approach~\cite{Berni2015}. We also note that by combining our strategy with a multi-pass metrology protocol~\cite{Higgins2007}, the sensitivity can be improved even further as in this case Heisenberg scaling will also apply to the number of measured samples~\cite{Borregaard2019}. The development and realization of a practical and loss-tolerant phase sensing scheme that beats the performance of any other current phase sensing strategy is not only of fundamental interest, but is also of practical relevance in phase sensing scenarios, where a low photon flux is needed to avoid the change of dynamics of the interrogated, potentially light-sensitive, sample~\cite{Landry2009,Waldchen2015}.  
\bibliographystyle{ieeetr}
\bibliography{Ref_PhaseEstimation}

\begin{thebibliography}{10}

\bibitem{Helstrom1969}
C.~W. Helstrom, ``{Quantum detection and estimation theory},'' {\em Journal of
  Statistical Physics}, vol.~1, no.~2, pp.~231--252, 1969.

\bibitem{Caves1981}
C.~Caves, ``{Quantum-mechanical noise in an interferometer},'' {\em Physical
  Review D}, vol.~23, no.~8, pp.~1693--1708, 1981.

\bibitem{Braunstein1994}
S.~L. Braunstein and C.~M. Caves, ``{Statistical distance and the geometry of
  quantum states},'' {\em Physical Review Letters}, vol.~72, no.~22,
  pp.~3439--3443, 1994.

\bibitem{Giovannetti2011}
V.~Giovannetti, S.~Lloyd, and L.~MacCone, ``{Advances in quantum metrology},''
  {\em Nature Photonics}, vol.~5, no.~4, pp.~222--229, 2011.

\bibitem{Demkowicz-Dobrzanski2015}
R.~Demkowicz-Dobrza{\'{n}}ski, M.~Jarzyna, and J.~Ko{\l}ody{\'{n}}ski,
  ``{Quantum Limits in Optical Interferometry},'' {\em Progress in Optics},
  vol.~60, pp.~345--435, 2015.

\bibitem{Polino2020}
E.~Polino, M.~Valeri, N.~Spagnolo, and F.~Sciarrino, ``Photonic quantum
  metrology,'' {\em AVS Quantum Science}, vol.~2, no.~2, p.~024703, 2020.

\bibitem{Dowling2008}
J.~P. Dowling, ``Quantum optical metrology – the lowdown on high-n00n
  states,'' {\em Contemporary Physics}, vol.~49, no.~2, pp.~125--143, 2008.

\bibitem{Mitchell2004}
M.~W. {Mitchell}, J.~S. {Lundeen}, and A.~M. {Steinberg}, ``{Super-resolving
  phase measurements with a multiphoton entangled state},'' {\em \nat},
  vol.~429, no.~6988, pp.~161--164, 2004.

\bibitem{Walther2004}
P.~{Walther}, J.-W. {Pan}, M.~{Aspelmeyer}, R.~{Ursin}, S.~{Gasparoni}, and
  A.~{Zeilinger}, ``{De Broglie wavelength of a non-local four-photon state},''
  {\em \nat}, vol.~429, no.~6988, pp.~158--161, 2004.

\bibitem{Nagata2007}
T.~Nagata, R.~Okamoto, J.~L. O'Brien, K.~Sasaki, and S.~Takeuchi, ``Beating the
  standard quantum limit with four-entangled photons,'' {\em Science},
  vol.~316, no.~5825, pp.~726--729, 2007.

\bibitem{Daryanoosh2018}
S.~Daryanoosh, S.~Slussarenko, D.~W. Berry, H.~M. Wiseman, and G.~J. Pryde,
  ``{Experimental optical phase measurement approaching the exact Heisenberg
  limit},'' {\em Nature Communications}, vol.~9, no.~1, pp.~1--6, 2018.

\bibitem{Slussarenko2017}
S.~Slussarenko, M.~Weston, H.~Chrzanowski, L.~Shalm, V.~Verma, S.~Nam, and
  G.~Pryde, ``Unconditional violation of the shot noise limit in photonic
  quantum metrology,'' {\em Nature Photonics}, vol.~11, 2017.

\bibitem{Gao2010}
W.-b. Gao, C.-y. Lu, X.-c. Yao, P.~Xu, O.~G{\"{u}}hne, A.~Goebel, Y.-a. Chen,
  C.-z. Peng, Z.-b. Chen, and J.-w. Pan, ``{Experimental demonstration of a
  hyper-entangled ten-qubit Schr{\"{o}}dinger cat state},'' {\em Nature
  Physics}, vol.~6, no.~March, pp.~331--335, 2010.

\bibitem{Wang2016}
X.-L. Wang, L.-K. Chen, W.~Li, H.-L. Huang, C.~Liu, C.~Chen, Y.-H. Luo, Z.-E.
  Su, D.~Wu, Z.-D. Li, H.~Lu, Y.~Hu, X.~Jiang, C.-Z. Peng, L.~Li, N.-L. Liu,
  Y.-A. Chen, C.-Y. Lu, and J.-W. Pan, ``Experimental ten-photon
  entanglement,'' {\em Phys. Rev. Lett.}, vol.~117, p.~210502, 2016.

\bibitem{Taylor2016}
M.~A. Taylor and W.~P. Bowen, ``{Quantum metrology and its application in
  biology},'' {\em Physics Reports}, vol.~615, pp.~1--59, 2016.

\bibitem{Monras2006}
A.~Monras, ``Optimal phase measurements with pure gaussian states,'' {\em Phys.
  Rev. A}, vol.~73, p.~033821, 2006.

\bibitem{Aspachs2009}
M.~Aspachs, J.~Calsamiglia, R.~Mu{\~{n}}oz-Tapia, and E.~Bagan, ``{Phase
  estimation for thermal Gaussian states},'' {\em Physical Review A - Atomic,
  Molecular, and Optical Physics}, vol.~79, no.~3, 2009.

\bibitem{Pinel2012}
O.~Pinel, J.~Fade, D.~Braun, P.~Jian, N.~Treps, and C.~Fabre, ``Ultimate
  sensitivity of precision measurements with intense gaussian quantum light: A
  multimodal approach,'' {\em Phys. Rev. A}, vol.~85, p.~010101, 2012.

\bibitem{Grangier1987}
P.~Grangier, R.~E. Slusher, B.~Yurke, and A.~LaPorta,
  ``Squeezed-light--enhanced polarization interferometer,'' {\em Phys. Rev.
  Lett.}, vol.~59, pp.~2153--2156, 1987.

\bibitem{Mckenzie2002}
K.~McKenzie, D.~A. Shaddock, D.~E. McClelland, B.~C. Buchler, and P.~K. Lam,
  ``Experimental demonstration of a squeezing-enhanced power-recycled michelson
  interferometer for gravitational wave detection,'' {\em Phys. Rev. Lett.},
  vol.~88, p.~231102, 2002.

\bibitem{LIGO2013}
{The LIGO Scientific Collaboration}, ``{Enhanced sensitivity of the LIGO
  gravitational wave detector by using squeezed states of light},'' {\em Nature
  Photonics}, vol.~7, p.~613, 2013.

\bibitem{Pradyumna2020}
S.~Pradyumna, E.~Losero, I.~Ruo-Berchera, P.~Traina, M.~Zucco, C.~Jacobsen,
  U.~Andersen, I.~Degiovanni, M.~Genovese, and T.~Gehring, ``Twin beam
  quantum-enhanced correlated interferometry for testing fundamental physics,''
  {\em Communications Physics}, vol.~3, no.~1, 2020.

\bibitem{Lawrie2019}
B.~J. Lawrie, P.~D. Lett, A.~M. Marino, and R.~C. Pooser, ``Quantum sensing
  with squeezed light,'' {\em ACS Photonics}, vol.~6, no.~6, pp.~1307--1318,
  2019.

\bibitem{Anisimov2010}
P.~M. Anisimov, G.~M. Raterman, A.~Chiruvelli, W.~N. Plick, S.~D. Huver,
  H.~Lee, and J.~P. Dowling, ``Quantum metrology with two-mode squeezed vacuum:
  Parity detection beats the heisenberg limit,'' {\em Phys. Rev. Lett.},
  vol.~104, p.~103602, 2010.

\bibitem{Berni2015}
A.~Berni, T.~Gehring, B.~Nielsen, V.~H{\"a}ndchen, M.~Paris, and U.~Andersen,
  ``Ab initio quantum-enhanced optical phase estimation using real-time
  feedback control,'' {\em Nature Photonics}, vol.~9, no.~9, pp.~577--582,
  2015.

\bibitem{Higgins2007}
B.~L. Higgins, D.~W. Berry, S.~D. Bartlett, H.~M. Wiseman, and G.~J. Pryde,
  ``{Entanglement-free Heisenberg-limited phase estimation},'' {\em Nature},
  vol.~450, no.~7168, pp.~393--396, 2007.

\bibitem{Borregaard2019}
J.~Borregaard, ``{Super sensitivity and super resolution with quantum
  teleportation},'' {\em npj Quantum Information}, no.~July 2018, pp.~1--6,
  2019.

\bibitem{Landry2009}
M.~P. Landry, P.~M. Mccall, Z.~Qi, and Y.~R. Chemla, ``{Characterization of
  Photoactivated Singlet Oxygen Damage in Single-Molecule Optical Trap
  Experiments},'' {\em Biophysj}, vol.~97, no.~8, pp.~2128--2136, 2009.

\bibitem{Waldchen2015}
S.~W{\"{a}}ldchen, J.~Lehmann, T.~Klein, S.~V.~D. Linde, and M.~Sauer,
  ``{Light-induced cell damage in live- cell super-resolution microscopy},''
  {\em Nature Publishing Group}, pp.~1--12, 2015.

\end{thebibliography}
\end{document}